\documentclass[prl,twocolumn,english,showpacs]{revtex4}
\usepackage{graphicx}
\usepackage{amssymb}
\usepackage{times}

\begin{document}

\title{Quantum transport through a Tonks-Girardeau gas}

\author{Stefan Palzer, Christoph Zipkes, Carlo Sias$^*$, and Michael K{\"o}hl}

\affiliation{Cavendish Laboratory, University of Cambridge, JJ Thompson Avenue, Cambridge CB3 0HE, United Kingdom}

\begin{abstract}
We investigate the propagation of spin impurity atoms through a strongly interacting one-dimensional Bose gas. The initially well localized impurities are accelerated by a constant force, very much analogous to electrons subject to a bias voltage, and propagate as a one-dimensional impurity spin wave packet. We follow the motion of the impurities in situ and characterize the interaction induced dynamics. We observe a very complex non-equilibrium dynamics, including the emergence of large density fluctuations in the remaining Bose gas, and multiple scattering events leading to dissipation of the impurity's motion.
\end{abstract}


\date{\today}

\maketitle
The impetus for miniaturization has resulted in the creation of nanostructures in which the motion of particles is purely one-dimensional. In these systems motional degrees of freedom can be excited only along one direction whereas in the two orthogonal directions the system occupies the quantum mechanical ground state. To reach the one-dimensional regime the chemical potential and the temperature need to be much smaller than the transverse level spacing. Interacting particles confined to a one-dimensional wave guide are fundamentally governed by many-body quantum mechanics \cite{Giamarchi2004}. In non-equilibrium situations this gives rise to genuine quantum dynamics, examples of which have been seen in single mode nanowires \cite{Auslaender2005} and in atom traps \cite{Moritz2003,Stoferle2004,Fertig2005,Kinoshita2006}. In this paper, we study the non-equilibrium transport of single or few impurity particles through a one-dimensional, strongly interacting Bose gas. The impurities are accelerated by a constant force, very much analogous to electrons subject to a bias voltage, and undergo scattering with the atoms in the Tonks-Girardeau gas.

An interacting one-dimensional Bose gas realizes a bosonic Luttinger liquid. Its many-body quantum state in the homogeneous case is characterized by a single parameter $\gamma=\frac{m g_{1D}} {\hbar^2 n_{1D}}$\cite{Lieb1963a,Lieb1963b}. Here $m$ is the atomic mass, $g_{1D}$ is the 1D coupling constant, and $n_{1D}$ is the 1D density. For weak interactions ($\gamma \ll 1$) Bose-Einstein condensation and superfluidity are possible in harmonically confined 1D systems. For strong interactions ($\gamma \gg 1$) the longitudinal motion of the particles is highly correlated. In this so-called Tonks-Girardeau regime the Bose gas "fermionizes", i.e. its $N$-particle wave function can be related to that of a $N$-particle spin-polarized Fermi gas \cite{Girardeau1960,Petrov2000b,Dunjko2001}. The density of the Bose gas as well as the density dependent correlation functions become Fermion-like and superfluidity vanishes \cite{Kheruntsyan2003,Caux2006}.

Both the weakly \cite{Moritz2003} and the strongly interacting \cite{Kinoshita2004,Paredes2004} regimes of one-dimensional Bose gases have been accessed a few years ago. This experimental realization of the one-dimensional Bose gas with $\delta$-functional interactions has triggered significant research efforts both experimentally and theoretically \cite{Bloch2008}. Of particular interest have been dynamical experiments \cite{Moritz2003,Stoferle2004,Fertig2005,Kinoshita2006}. Elementary transport experiments have seen the suppression of dipole oscillations in a corrugated potential \cite{Fertig2005} and the absence of thermalization \cite{Kinoshita2006}. These experiments, however, have focused on global properties of the gas rather than using single impurity probes. Moreover, in contrast to previous non-equilibrium experiments in one dimension, we work with an open quantum system in which the impurity atoms continuously gain kinetic energy and can transfer this energy into the trapped gas by collisions. On the theoretical side, single particle perturbations have been studied in a number of different regimes \cite{Astrakharchik2004b,Fuchs2005,Kollath2005b,Zvonarev2007,Kleine2008,Girardeau2009,Gangardt2009,Lamacraft2009}.

\begin{figure}[htbp]
  \includegraphics[width=\columnwidth,clip=true]{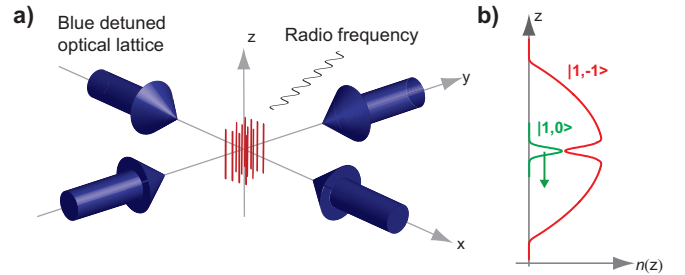}
  \caption{(Color online) {\bf a)} An array of one-dimensional Bose gases in a two-dimensional blue-detuned optical lattice. Vertical confinement is provided by a harmonic magnetic potential $B(z)$. {\bf b)} Creating impurities by a radio frequency pulse resonant at a specific magnetic field. The impurity atoms in the $|F=1, m_F=0\rangle$ state experience no vertical confining potential and are accelerated by gravity into the -z-- direction.}
  \label{fig1}
\end{figure}
\begin{figure*}[htbp]
\includegraphics[width=0.8\textwidth]{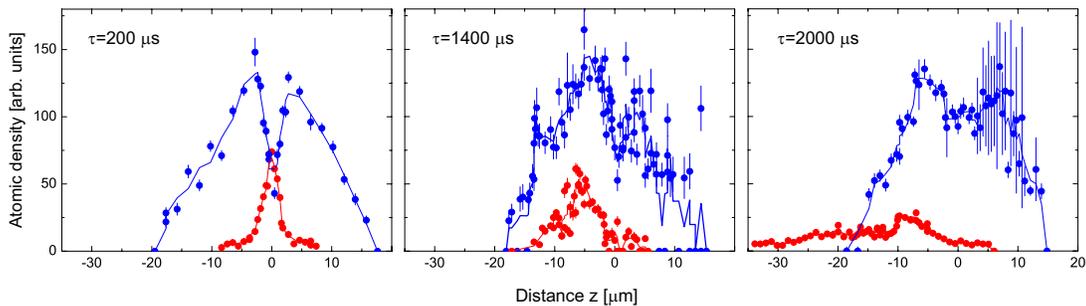}
  \caption{(Color online) In situ measurement of the time evolution of both the trapped component (blue) and the impurity (red) for different times $\tau$. The data are taken for $\gamma=7$. The solid line is a two-point average of the data to guide the eye. }
  \label{fig2}
\end{figure*}

Our realization of strongly interacting one-dimensional gases is depicted in Figure \ref{fig1}a. We start from producing almost pure Bose-Einstein condensates of $^{87}$Rb of up to $1.5\times 10^5$ atoms in a magnetic trap in the hyperfine ground state $|F=1,m_F=-1\rangle$. The harmonic magnetic trap has the frequencies $\omega_{x,z}=2 \pi \times 39$\,Hz and $\omega_y=2 \pi \times 11$\,Hz. Strong confinement into a one-dimensional geometry is achieved by adiabatically loading the three-dimensional Bose-Einstein condensate into an optical lattice. The optical lattice is formed by two retro-reflected laser beams of wavelength $\lambda=764$\,nm arranged in the horizontal xy--plane. In this blue-detuned optical lattice the atoms are trapped in the intensity minima of the interference pattern. Thus the vertical confinement is purely magnetic. At the position of the condensate, the standing wave laser fields overlap perpendicularly with orthogonal polarizations and are focused to a circular waist (1/$e^2$-radius) of 180 $\mu$m. The frequencies of the two beams are offset with respect to each other by 280\,MHz.
The optical potential depth $U$ is proportional to the laser intensity and can be expressed in terms of the recoil energy $E_{rec}=\frac{h^2}{2 m \lambda^2}$. Adiabatic loading into the ground state of the optical lattice was achieved by ramping up the laser intensity with a linear ramp of 150\,ms duration. Typically, we confine approximately 50 atoms per one-dimensional tube in the center of the three-dimensional gas cloud. The interaction parameter $\gamma$ depends on the strength of optical lattice and ranges between 2 and 10. Here $\gamma$ represents the density-weighted average of $\gamma_\textrm{tube}$ across all tubes. The error in the determination of $\gamma$ is $15\%$ dominated by uncertainties in atom number and calibration of the lattice depth.

The hybrid magnetic/optical trapping potential provides us with the necessary means to create and detect the impurities with very good spatial resolution. We create spin impurities by using spatially resolved radio frequency manipulation. We apply a pulse (200\,$\mu$s) of radio frequency resonant with the $|F=1, m_F=-1\rangle \rightarrow |F=1, m_F=0\rangle$ transition \cite{Bloch1999}. The spatial width of our addressing region is Fourier-limited by the duration of the pulse and is $\Delta z \approx 2.3\mu$m (see Figure \ref{fig1}b). We drive approximately a $\pi/2$ pulse producing an impurity wave packet containing up to 3 atoms per one-dimensional tube. The atoms in the impurity state have zero magnetic moment and are accelerated downwards by gravity. They start from zero center-of-mass velocity and reach several times the velocity of sound at the edge of the trapped cloud. During the propagation they are continuously confined to the one-dimensional waveguide in the radial direction because the kinetic energy acquired is less than the radial level spacing. On their way downwards they strongly interact with the remaining atoms because the three-dimensional scattering lengths for collisions in the three possible combinations of states are approximately equal to each other.

In order to accurately study the motion of the impurity atoms in our one-dimensional Bose gas we have performed a time resolved tomographic measurement of both the Bose gas and the impurity density distribution. At a variable time $\tau$ after preparation of the impurity wave packet we measure the density distribution in each of the components separately in situ. We employ magnetic field (position) sensitive microwave transitions (pulse duration 200 \,$\mu$s) between the hyperfine ground states $|F=1,m_F=-1\rangle \rightarrow |F=2,m_F=0\rangle$ (trapped gas) and $|F=1,m_F=0\rangle \rightarrow |F=2,m_F=1\rangle$ (impurity) and detect the atoms by absorption imaging on the $|F=2\rangle \rightarrow |F'=3\rangle$ transition of the $D_2$ line. Figure \ref{fig2} shows three snapshots of the time evolution of both the trapped component (blue) and the impurity (red). We observe that the impurity atoms are initially  well localized in a compact wave packet whose width is in agreement with the Fourier limit. As the impurity wave packet propagates it becomes wider and distorted indicating a strong dispersion and dissipation. For example, at ($\tau$ =2000 $\mu$s, z=-10 $\mu$m) we observe a very distinct asymmetric steepening of the propagating impurity wave packet which resembles the proposed shape of a supersonic shock wave in one dimension \cite{Damski2004}. Moreover, the propagation of the impurity leaves a strong imprint on the trapped component indicating a strong mutual interaction. We observe that the statistical noise of the data points in the sample after propagation of the impurity ($\tau$ =1400 $\mu$s and $\tau$ =2000 $\mu$s) is considerably larger than for the initial state. This could be attributed to high frequency oscillations of the atomic density at a length scale below our spatial resolution limit.

\begin{figure}[htbp]
\includegraphics[width=.75\columnwidth]{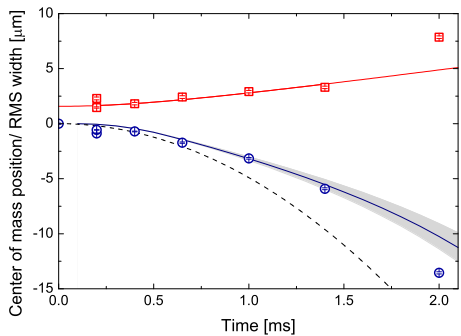}
  \caption{(Color online) The circles show the measured center-of-mass position 
  taken for $\gamma=7$. The error bars are the statistical error of the center of mass of the measured density distribution. The solid line is the prediction according to the model described in the text. The gray shaded area indicates the regime of uncertainty of $10\%$ of $n_{1D}$ given by our experimental parameters. The dashed curve indicates purely ballistic motion. The squares show the increase of the width of the impurity wave packet. The data point at 2\,ms contains atoms which have already left the trapped gas which is not taken into account by the theory.}
  \label{fig3}
\end{figure}

We have analyzed the time evolution of the center-of-mass and the width of the impurity component (see Figure \ref{fig3}). Very clearly, the center-of-mass of the impurity does not follow a ballistic trajectory but its motion is hindered by the presence of the interacting Bose gas. Theoretical investigations for impurities of equal mass exist only for the case of non-accelerated impurities \cite{Fuchs2005,Kollath2005b,Zvonarev2007,Kleine2008,Girardeau2009,Gangardt2009,Lamacraft2009} and mostly for low momenta of the impurity. In our case, however, the impurity atoms are accelerated by gravity and reach the velocity of sound $c$ already after traveling a distance shorter than the interparticle separation $1/n_{1D}$. In the conceptually simplest case of binary collisions the probability for a momentum change in a collision between two atoms in 1D (a reflection of the impurity particle from the majority atom) is $p=(1+(a_{1D} m v/\hbar)^2)^{-1}$ \cite{Olshanii1998}. Here $a_{1D}=2\hbar^2/(m g_{1D})$ is the one-dimensional scattering length. For the first collision (i.e. after traveling a distance of order $1/n_{1D}$) we find $p\approx 0.2$ which becomes quadratically smaller for subsequent collisions as the velocity of the accelerated impurity particle increases. Consequently, in the majority of the binary collisions the impurity is transmitted. For binary collisions the energy dissipation scales like $\dot{E}_{bin}(v)\propto v^3/(1+ (m a_{1D} v/\hbar)^2)$ \cite{Olshanii1998,Kinoshita2006}. Refining this result for a many-body system we adopt an approach based on Fermi's golden rule. The rate of scattering $\Gamma(k_i,k_f)$ a single impurity particle from an initial momentum state $k_i$ to a final momentum state $k_f$ can then be determined by the dynamic structure factor $S(q,\omega)$ of the gas: $\Gamma(k_i,k_f)\propto \int d\omega S(q,\omega)$ \cite{Timmermans1998}. Here $q=k_i-k_f$ is the momentum change of the impurity, $\hbar \omega=\epsilon(k_i)-\epsilon(k_f)$ is its corresponding energy loss, and $\omega_q$ is the excitation spectrum of the gas. We assume a free particle like dispersion relation for the impurity atoms (which is viable to first order in the impurity--gas interaction strength \cite{Lamacraft2009}). For a homogeneous weakly interacting system we have calculated the energy dissipation rate of the impurity using $\dot{E}(k_i)\propto \int dq d\omega \omega S(q,\omega)$ \cite{Timmermans1998} to be $\dot{E}_{SF}(v)= \frac{2 \hbar^2 n_{1D}}{m a_{1D}^2} v \left(1-\left(\frac{c}{v}\right)^4\right)$ for $v>c$. In a homogeneous Tonks gas we find the energy dissipation $\dot{E}_{TG}(v)= \frac{2 \hbar^2 n_{1D}}{m a_{1D}^2} v=- \alpha m v$ for $v>v_F=\hbar \pi n_{1D}/m$, i.e. a constant, density dependent force. In the limit of large velocities both results agree with the result for binary collisions. In figure \ref{fig3} we plot the predicted trajectory from a numerical integration of the equation of motion taking into account a Thomas-Fermi profile of the density of the majority component as well as antibunching over a length scale of $1/n_{1D}$. We use the predicted value of the force constant $\alpha(z)=-\frac{2 \hbar^2 n_{1D}(z)}{m^2 a_{1D}^2}$ with $\alpha(z=0)= -8\,$m/s$^2$ and find very good agreement with the experimental data without any adjustable parameters. This theoretical model neglects the distortion of the density of the trapped gas as well as coherent or multiple scattering processes and back action onto the impurity atoms.

The impurity wave packet spreads considerably as it propagates. We have measured the root-mean-square (rms) radius $w$ of the wave packet and plot its evolution in Figure \ref{fig3}. A function $w(t)= \sqrt{w_0^2+v^2t^2}$ is fitted to the data to extract the velocity $v$ with which the wave packet spreads while it is fully inside the trapped gas. We find $v=1.5\pm 0.4$\,mm/s which is slightly smaller than the uncertainty limited momentum distribution of the impurity wave packet of 2.0\,mm/s.

\begin{figure}[htbp]
  \includegraphics[width=.8\columnwidth,clip=true]{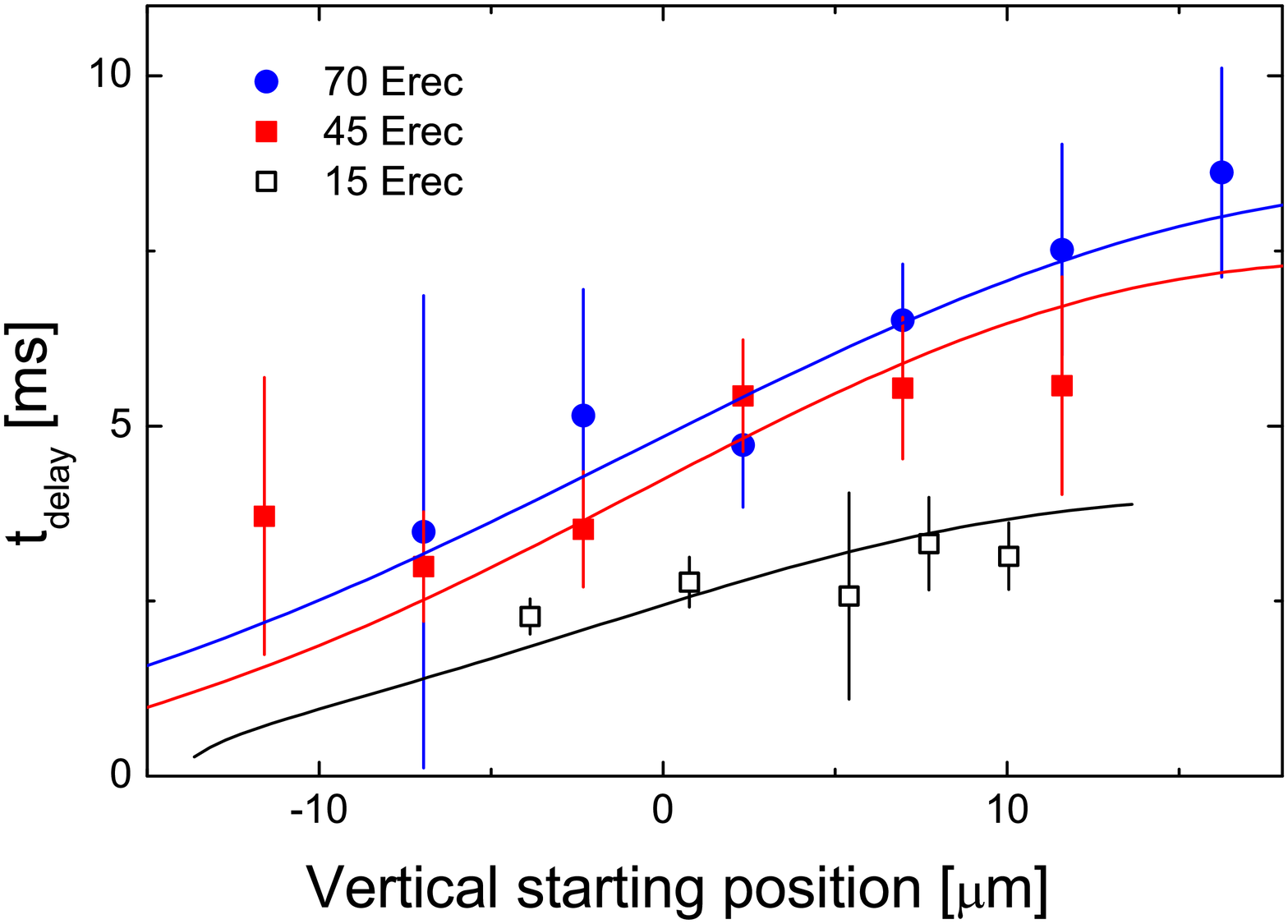}
  \caption{(Color online) Time delay between the impurity preparation and the exit of the last impurity atoms from the trapped cloud for different lattice depths. The starting position is relative to the center of the trapped cloud. The solid lines show fits according to the model described in the text using a Thomas-Fermi profile for the density. The numerical prefactor agrees with the model to within a factor of 3.}
  \label{fig4}
\end{figure}
We have investigated the number of collisions the impurities are undergoing during the transport and the time lag of their motion. Using state-selective detection we determine how the number of impurity atoms inside the trapped component decays as a function of time. From these data we extrapolate the time delay after which all impurity atoms have left the trapped Bose gas for different initial positions and for different values of $\gamma$ (see Figure \ref{fig4}). In a simple model we assume that in every collision event the impurity motion is set back to zero velocity and afterwards the atoms are accelerated again by gravity. On average the time between two collision events can be estimated $t_{coll}\approx \sqrt{2/(n_{1D}(z)g)}$ in which $g$ is the gravitational acceleration. The delay time as compared to ballistic motion can be expressed as the total number of inelastic collisions times the time delay accumulated per collision event (approximately $t_{coll}$) resulting in $\int_{-R}^{z_0} n_{1D}(z) \Gamma(z) t_{coll}(z)^2\,dz$. Here $\Gamma(z)=\frac{4 \hbar^2 n_{1D}(z)}{m^2 a_{1D}^2 v}$ is the collision rate for $v\approx\sqrt{2g/n_{1D}(z)}>c$. This simple model explains the observed behavior well (see fits in Figure \ref{fig4}). The number of collisions is on the order of $5\lesssim t_{delay}/t_{coll}\lesssim 20$, depending on the interaction strength and the starting position.

\begin{figure}[htbp]
  \includegraphics[width=.8\columnwidth,clip=true]{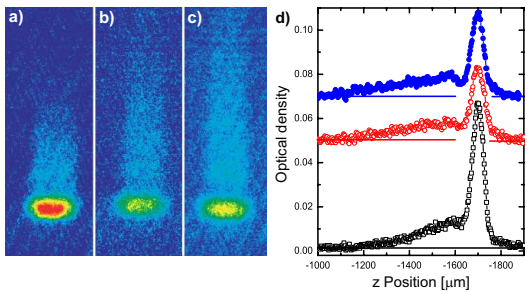}
  \caption{(Color online) Time of flight image of the impurity atoms and corresponding profiles of the optical column density. The images are averaged over 5-10 repetitions of the experiment. Impurities released from a lattice of potential depth $15\,E_{rec} (\gamma=3)$ (a, squares), $30\,E_{rec} (\gamma=5)$ (b, open circles), and $45\,E_{rec} (\gamma=7)$ (c, full circles) taken 18.7\,ms after the radio frequency pulse. Both the number of atoms scattered out of the main peak and the length of the distribution of the scattered atoms increases with increasing $\gamma$. The curves in (d) have been offset vertically for clarity. The width of the atomic density distribution corresponds to the first Brillouin zone of the optical lattice.}
  \label{fig5}
\end{figure}

Figure \ref{fig5} shows absorption images of impurities propagated out of a Tonks gas for different values of $\gamma$ while the one-dimensional confinement was kept on.
The image is recorded a fixed time after the preparation of the impurity wave packet. Impurity atoms leave the Tonks-Girardeau gas after a variable time depending on how many collisions they undergo. After the atoms have left the sample they simply undergo ballistic motion. We observe a leading wave packet followed by a relatively long tail of atoms. From the position of the main peak we conclude that the atoms have not undergone momentum changing scattering while moving through the one-dimensional sample. The width of the main peak corresponds to an rms velocity spread of $(1.9 \pm 0.1)$\,mm/s, in agreement with the uncertainty limited width of the impurity wave packet. The atoms in the tail are scattered out of the main peak and, qualitatively, the tail represents the one-dimensional analog of the s-wave scattering spheres previously observed in three dimensions \cite{Chikkatur2000} but including multiple scattering events. The length of the tail is in good agreement with the results presented in Figure \ref{fig4}. We observe a distinct minimum between the main peak and the tail (see Figure \ref{fig5}d) for large values of $\gamma$. This minimum could reflect the increasing fermionic nature of the atoms in the gas. Since our impurity initially was a constituent of the Tonks-Girardeau gas, the two-body collision rate may still be related to the density-density correlation function $g^{(2)}(r)$ and becomes suppressed on the length scale of $r\approx 1/n_{1D}$ as the gas fermionizes \cite{Kheruntsyan2003}. Then collisions between the impurities and the trapped atoms would be suppressed at the very beginning of the dynamics. Quantitatively, the width of the minimum of scattered atoms in the time-of-flight image agrees with a suppression of collisions over the length scale $\approx 1/n_{1D}$ in situ. A comparison experiment performed with a weakly interacting three-dimensional Bose-Einstein condensate did not reveal a similar feature.

In conclusion, we have studied quantum transport of spin impurity atoms through a strongly interacting one-dimensional Bose gas. 
For pseudospin-1/2 or spin-1 Bose gases as the majority component our tomographic detection technique could reveal fundamental properties of spin transport \cite{Zvonarev2007} and spin-charge separation \cite{Kleine2008}.

We are grateful to D. Gangardt, C. Kollath, B. Simons, and W. Zwerger for discussions and the workshop of the Cavendish Laboratory, M. Goodrick, and T. St{\"o}ferle for support. We acknowledge support from {EPSRC} (EP/F016379/1) and the Herchel Smith Fund (CS).

\end{document}